\documentclass[final,5p,times,twocolumn]{elsarticle}
\usepackage{ifpdf}
\usepackage[usenames]{color}
\usepackage{graphicx}
\usepackage{amssymb}
\usepackage{amsmath}
\usepackage{ulem}
\journal{\textbf{Chaos, Solitons and Fractals}}
\begin{document}

\begin{frontmatter}

\title{Dynamical stabilization of two-dimensional trapless Bose-Einstein condensates by three-body interaction and quantum fluctuations}

\author[pu,bdu]{S. Sabari}
\author[pu2]{K. Porsezian}
\author[bdu]{P. Muruganandam}
\address[pu]{Department of Physics, S.P. Pune University, Pune, 411007, India}
\address[pu2]{Department of Physics, Pondicherry University, Puducherry 605014, India}
\address[bdu]{Department of Physics, Bharathidasan University, Palkaliperur Campus, Tiruchirappalli 620024, India}

\date{\today}

\begin{abstract}
Analyzing a Gross-Pitaevskii equation with cubic, quartic, and quintic nonlinearities through analytical and numerical methods, we examine the stability of two-dimensional (2D) trapless Bose-Einstein condensates (BECs) with two-, three-body interactions and quantum fluctuations. Applying a variational approach, we derive the equation of motion and effective potential to discuss in detail the stability of the BECs in 2D free space. We show that with the aid of quantum fluctuations it is possible to stabilize 2D trapless BEC without any oscillatory nonlinearities. Also, there is an enhancement of the stability of the system, due to the inclusion of the three-body interaction and quantum fluctuations in addition to the two-body interaction. We further study the stability of 2D trapless BECs with rapid periodic temporal modulation of scattering length by using a Feshbach resonance. We discuss all possible ways of stabilization of trapless BECs in 2D by three-body interaction and quantum fluctuations. Finally, we verify our analytical results with numerical simulation using split-step Crank-Nicholson method. These match well with the analytical predictions.
\end{abstract}

\begin{keyword}
 Bose-Einstein condensate, Gross-Pitaevskii equation, quantum fluctuation, Crank-Nicholson method
\PACS 03.75.-b, 05.45.Yv, 42.65.Tg, 42.50.Lc
\end{keyword}

\end{frontmatter}

E-mail: ssabari01@gmail.com

\section{Introduction}
\label{sec1}

\qquad After a successful experimental realization of Bose-Einstein condensates (BECs)~\cite{Anderson1995,Davis1995,Bradley1995Bradley1997}, there have been numerous studies on the dynamics and properties of the condensates. BECs at very low temperatures are usually described by the nonlinear, mean-field Gross-Pitaevskii (GP) equation~\cite{Dalfovo1999}. The effect of the inter-atomic interaction is accounted for by a nonlinear term in the GP equation. The s-wave scattering length, $a_s(t)$, plays an important role in the description of atom-atom interaction at ultralow temperatures. The magnitude and sign of the s-wave scattering length, $a_s(t)$, can be tuned to any value, large or small, positive or negative by applying an external magnetic field. It is given by $a_s(t) = a \left[ 1 + \Delta/\left( B_0 - B(t) \right) \right]$, where $B(t)$ is the externally applied magnetic field, $\Delta$ is the width of resonance and $B_0$ is the resonant value of the magnetic field.

It is understood that at low density, where the inter-atomic distances are much greater than the distance scale of atom-atom interactions, two-body interaction can be described by a scattering length where the effects of the higher-order interactions are negligible~\cite{Dalfovo1999,Gammal2000}. But, in some experiments, the density of the atoms is considerably high. Consequently, the simple mean-field GP equation with two-body interaction alone becomes less valid. Hence, the dynamics of the BEC need a better description of the atom-atom interaction. Such a system comprises three-body interactions due to higher densities and quantum fluctuations due to a lower-order correction to the two-body collision potential~\cite{TDLEE}. This system is governed by a GP equation with cubic, quartic, and quintic nonlinearities~\cite{QFrecent}. In the present study, we aim to derive a modified GP equation which goes beyond the mean-field theory and to discuss the stability of the Bose-Einstein condensates at high density in the presence of quantum fluctuations~\cite{QFrecent,qfDBEC}.

Moreover, it is well known that for an attractive interaction~\cite{Bradley1995Bradley1997}, the condensate is stable only upto a maximum critical number of atoms. When the number of atoms increases beyond this critical value, due to the inter-atomic attraction, the radius of the BEC tends to zero and the maximum density of the condensate tends to infinity~\cite{Adhikari2001}. With a supply of atoms from an external source the condensate can grow again and thus a series of collapses can take place, this has been observed experimentally in BECs of $^7${Li} with attractive interaction~\cite{Bradley1995Bradley1997}. Theoretical analysis based on the GP equation also confirms the collapse. Thus for a system of atoms with attractive two-body interaction, the condensate has no stable solution above a certain critical number of atoms ($N_{max}$)~\cite{Edwards1995,Adhikari2000,Ruprecht1995,Sabariepjd}. However, as reported by Gammal et al~\cite{Gammal2000}, the addition of a repulsive potential derived from three-body interaction can play an important role in the description of the static and the dynamic properties of condensates and is consistent with a number of atoms larger than $N_{max}$. Therefore, inclusion of effects of the three-body interactions in the governing equation of condensates is necessary.

Further, in 2D or 3D free space, any attempt to create a soliton leads to either collapse or spreading such that no stabilization of the solitary wave can be realized~\cite{strecker}. However, a scheme for the soliton stabilization in 2D free space is suggested by a rapid periodic temporal modulation of the two-body interaction~\cite{Adhikari2004,Abdullaev2003,Saito2003}. Also, in recent years, studies of temporal and spatially modulated nonlinearities have attracted considerable attention in several areas, for example, nonlinear physics~\cite{BAM_RMP,Kivshar2003}, optics~\cite{BAM_RMP,Kivshar2003,RBABU,Towers2002,Zeng12,Adhikari04,Adhikari05,Dai11} and BECs~\cite{Adhikari2004,Abdullaev2003,Saito2003,BAM_RMP,Sabari2010,Sabari2015,Liu10,adhikari2003,JHE}. The time-independent GP equation yields only the solution of stationary problems. In the present study, through the time-dependent GP equation, we reexamine the problem of stabilization and point out the inclusion of the three-body interaction and quantum fluctuations can lead to a stabilization of the trapless 2D BECs at higher-densities. Likewise, the temporal modification of the scattering length can also increase the stabilization of the trapless system. In this paper, in addition to analytical studies, we also perform numerical verification for the stability of trapless BECs in the presence of constant and oscillatory three-body interactions and quantum fluctuations. Our present analysis strongly suggests that the inclusion of three-body interaction and quantum fluctuations of suitable form can stabilize the trapless BECs in 2D. We further illustrate from numerical simulations that the trapless condensate can maintain a reasonably constant spatial profile over a sufficient interval of time through temporal modulation. The organization of the present paper is as follows. In Sec.~\ref{sec2}, we derive a modified GP equation and discuss the variational study of the problem and point out the possible stabilization of a trapless 2D BECs with three-body interaction and quantum fluctuations. In Sec.~\ref{sec3}, we report the numerical results of the time-dependent GP equation with two- and three-body interactions and in the presence of quantum fluctuations, through the split-step Crank-Nicholson (SSCN) method. Finally, we give the concluding remarks in Sec.~\ref{sec4}.

\section{The model and Variational approximation}
\label{sec2}
We consider Bose-Einstein condensates with two-, three-body interactions and quantum fluctuations. At ultra-low temperatures, this system can be described by the following dimensionless, time-dependent Gross-Pitaevskii equation with cubic, quartic, and quintic nonlinearities~\cite{TDLEE,QFrecent,MGPE}:
\begin{align}
\mathrm{i}&\hbar\frac{\partial \psi(\mathbf{r},t)}{\partial t} =  \Big[-\frac{\hbar^2}{2m}\nabla^2  +  \frac{1}{2}m\omega^2   {r}^2 +g(t)_{2b} N\vert \psi(\mathbf{r},t)\vert ^2 \notag \\ & 
+ q(t)_{qf} N^{3/2}\vert \psi(\mathbf{r},t)\vert ^3 +\chi(t)_{3b} N^2\vert \psi(\mathbf{r},t)\vert ^4 \Big]
\psi(\mathbf{r},t), \label{Jac1}
\end{align} 
where $\hbar$ is the reduced Planck's constant, $m$ is the mass of the boson, $N$ is the number of atoms in the condensate. The coefficient $g(t)_{2b}$ is the strength of the two-body interaction which can be tuned to any desired value by using Feshbach resonance technique and $q(t)_{qf} $ is the Lee-Huang-Yang term that accounts for the correction due to quantum fluctuations~\cite{TDLEE}, and $\chi(t)_{3b} $ is the strengths of the three-body interactions. The normalization condition is $\int \vert \psi(\mathbf{r},t)\vert ^2 d  {r}  = 1$.

It is more convenient to use the GP equation (\ref{Jac1}) in  a dimensionless form. For this purpose we shall make the transformation of variables as $r' =   {r}/l$, $t' = t\omega$, $l = \sqrt{\hbar/(m\omega)}$ and  $\phi(r',t')=\psi(  {r},t)(l^3/2\sqrt{2})^{1/2}$. Considering the ground state with zero angular momentum~\cite{Adhikari2001,Adhikari2000,zero_angu}, the radial part of the GP equation (\ref{Jac1}), after dropping the primes, can be written as~\cite{TDLEE,QFrecent,MGPE},
\begin{align}
\mathrm{i}\frac{\partial \phi(r,t)}{\partial t}= & \Bigg[- \frac{1}{2}\nabla_{2D}^2+ \frac{1}{2} d(t) r^2+ g(t) \left\vert\phi(r,t)\right\vert^2\notag\\
& + q(t) \left\vert\phi(r,t)\right\vert^3+\chi(t)  \left\vert\phi(r,t)\right\vert^4 \Bigg]\phi(r,t),  \label{Jac2}
\end{align}
where $r^2=x^2+y^2$, 
$d(t)$ represents the strength of the external trap, which is to be reduced from $1$ to $0$ when the trap is switched off. Usually the strength of the three-body interaction is very small when compared with strength of the two-body interaction~\cite{Gammal2000}. But, in some cases where the three-body interaction is dominant, it can be roughly of the same magnitude and range as the two-body term~\cite{Brunner2004}. In some other regimes, the three-body interaction can even become more dominant, for instance, the cases of higher densities and Efimov resonance~\cite{Bulgac2004}. Recent works have suggested the possibility of controlling the strength of the three-body interaction between atoms independent of the control of the two-body collisions~\cite{Buchler2007Paredes2007}. These findings have triggered numerous works on BECs with two- and three-body interactions~\cite{Gammal2000,Sabari2010,Abdullaev2001,Wamba2008,Ping2009,Ercy1996,Josser1997,Fetter1971}.
In the following, we use the variational approach with the trial wave function (Gaussian ansatz) for the solution of  (\ref{Jac2})~\cite{VAM}:
\begin{align}
\phi(r,t)=N(t) \exp{\left[-\frac{r^2}{2\,R(t)^2}+\frac{\mathrm{i}}{2} \,\beta(t)\,r^2 \right]}, \label{Jac3}
\end{align}
where $N(t) = 1/ \sqrt{\pi}R(t) $, $R(t)$ and $\beta(t)$  are the normalization, width and chirp, respectively. The Lagrangian density for Eq.~(\ref{Jac2}) with $d(t) = 0$ is given by
\begin{align}
\mathcal{L} = & \frac{\mathrm{i}}{2}\left(\frac{\partial\phi}{\partial t}\phi^*-\frac{\partial\phi^*}{\partial t}\phi\right) -\frac{1}{2}\left\vert \frac{\partial\phi}{\partial r}\right\vert ^2  - \frac{1}{2}g(t)\vert \phi\vert ^4  
\notag\\ 
& -\frac{2 }{5}q(t)\vert \phi\vert ^5 -\frac{1}{3}\chi(t) \vert \phi\vert ^6 . \label{Jac4}
\end{align} 
The trial wave function (\ref{Jac3}) is substituted in the above Lagrangian density (\ref{Jac4})  and an effective Lagrangian is calculated as $L_{eff} = 2\pi\int_0^{\infty} \mathcal{L}(\phi) \,r  dr$. 
The following Euler-Lagrangian equations are then obtained from the effective Lagrangian.
\begin{align}
\frac{d R}{dt} =  &\, R(t) \, \beta(t),\label{Jac6} \\ 
\frac{d \beta}{dt}   = &\, \frac {1} {R(t)^4} 
  + \frac{g(t)}{2\pi R(t)^4} +\frac{12q(t)}{25\sqrt{\pi^3} R(t)^5}
  +\frac{4\chi(t)}{9 \pi^2 R(t)^6} - \beta(t)^2 .\label{Jac7}
\end{align}
By combining the Eqs.~(\ref{Jac6}) and (\ref{Jac7}), we get the following second-order differential equation for the evolution of the width,
\begin{align}
\frac{d^2 R}{dt^2}  = & \frac{1}{R(t)^3}+ \frac{g_0 + g_1 \sin \Omega t }{2 \pi R(t)^3} +\frac{12 ( q_0 + q_1 \sin \Omega t ) }{25\pi^{3/2} R(t)^4}  \notag\\ & +\frac{4 ( \chi_0 + \chi_1 \sin \Omega t ) }{9 \pi^2 R(t)^5}, \label{Jac8}
\end{align}
with $ g(t) = g_0 + g_1 \sin \Omega t$, $q(t)=q_0 + q_1  \sin \Omega t  $ and $\chi(t) =  \chi_0 +\chi_1  \sin \Omega t  $, where $g_0 (g_1)$, $q_0 (q_1)$ and $\chi_0 (\chi_1)$ are the constant (oscillating) part of the two-body, quantum fluctuations, and three-body interaction, respectively. Now $R(t)$ can be separated into a slowly varying part $R_0(t)$ and a rapidly varying part $B(t)$ by $R(t)=R_0(t)+B(t)$. When $ \Omega\ \gg 1 $, $B(t)$ becomes of the order of $\Omega^{-2}$. Keeping the terms of the order of up to $\Omega^{-2}$ in $B(t)$, one may obtain the following equations of motion for $R_0(t)$ and $B(t)$~\cite{Landau1960},
\begin{align}
\frac{d^2 B}{dt^2}  = & \frac{g_1 \sin \Omega t}{2 \pi R_0(t)^3}+\frac{12 q_1  \sin \Omega t   }{25\pi^{3/2} R_0(t)^4} +\frac{4 \chi_1  \sin \Omega t  }{9 \pi^2 R_0(t)^5}, \label{Jac9} \\
\frac{d^2 R_0}{dt^2} = & \frac{1}{R_0(t)^3}+ \frac{g_0 }{2 \pi R_0(t)^3}+\frac{12 q_0 }{25\pi^{3/2} R_0(t)^4} +\frac{4\chi_0 }{9 \pi^2 R_0(t)^5}\notag\\
& -\frac{3g_1\left\langle{B(t)  \sin \Omega t  }\right\rangle}{2 \pi R_0(t)^4}-\frac{48 q_1\left\langle{B(t)  \sin \Omega t  }\right\rangle}{25\pi^{3/2} R_0(t)^5} \notag\\  
& -\frac{20\chi_1\left\langle{B(t)  \sin \Omega t  }\right\rangle}{9 \pi^2 R_0(t)^6},\label{Jac10}
\end{align}
where the $\langle \cdots \rangle$ indicates the time average of the rapid oscillation. From Eq.~(\ref{Jac9}) we can get,
\begin{align}
B(t)= - \frac{g_1 \sin \Omega t }{2 \pi \Omega^2 R_0(t)^3}  - \frac{12 q_1 \sin  \Omega t }{25\pi^{3/2} \Omega^2 R_0(t)^4} - \frac{4 \chi_1 \sin \Omega t }{ 9 \pi^2 \Omega^2 R_0(t)^5 },
\end{align}
and substituting $B(t)$ into Eq.~ (\ref{Jac10}), we obtain the following equation of motion for the slowly varying part,
\begin{align}
\frac{d^2 R_0}{dt^2} = & \frac{1}{R_0^3}+ \frac{g_0 }{2\pi R_0^3}+\frac{12 q_0 }{25\pi^{3/2} R_0^4} +\frac{4\chi_0 }{9 \pi^2 R_0^5}+ \frac{3g_1^2} {8\pi^2 \Omega^2 R_0^7 } \notag\\ & 
+\frac{21g_1 q_1} {25 \pi^{5/2} \Omega^2 R_0^8}+\frac{288 q_1^2}{625\pi^3 \Omega^2 R_0^9}+ \frac{8g_1 \chi_1} {9 \pi^3 \Omega^2 R_0^9 } \notag\\ & +\frac{24 q_1 \chi_1} {25 \pi^{7/2} \Omega^2 R_0^{10}}+ \frac{40\chi_1^2} {81 \pi^4 \Omega^2 R_0^{11}},\label{Jac11}
\end{align}
and the effective potential $U(R_0)$ corresponding to the above Eq.~(\ref{Jac11}) can be written as,
\begin{align}
U(R_0) = & 
\frac{1}{2R_0^2}+ \frac{g_0 }{4\pi R_0^2}+\frac{4 q_0 }{25\pi^{3/2} R_0^3} 
   + \frac{\chi_0 }{9 \pi^2 R_0^4}  \notag\\ & 
+ \frac{g_1^2} {16\pi^2 \Omega^2 R_0^6} +\frac{3 g_1 q_1} {25 \pi^{5/2} \Omega^2 R_0^7}
   + \frac{36 q_1^2}{625\pi^3 \Omega^2 R_0^8}  \notag\\ & 
+ \frac{g_1 \chi_1} {9 \pi^3 \Omega^2 R_0^8}+\frac{8 q_1 \chi_1} {75 \pi^{7/2} \Omega^2 R_0^{9}}
   + \frac{4\chi_1^2} {81 \pi^4 \Omega^2 R_0^{10}}, \label{eq:veff:0}
\end{align}
Now from the nature of the effective potential, we can investigate the stability of the system in the presence and the absence of quantum fluctuations and three-body interaction. It may be noted that one needs a minimum attraction to have a stable system in free space, otherwise it will expand. Small oscillations around a stable configuration are possible when there is a minimum in this effective potential~\cite{Adhikari2004,Saito2003}. 

\subsection{Effect of quantum fluctuation on the stability of BECs with two-body interaction}

In the absence of three-body interaction, the effective potential can be written as,
\begin{align}
U(R_0) = & \frac{1}{2R_0^2}+ \frac{g_0 }{4\pi R_0^2}+\frac{4 q_0 }{25\pi^{3/2} R_0^3} + \frac{g_1^2} {16\pi^2 \Omega^2 R_0^6}
\notag\\ & 
+\frac{3 g_1 q_1} {25 \pi^{5/2} \Omega^2 R_0^7}+\frac{36 q_1^2}{625\pi^3 \Omega^2 R_0^8}, \label{eq:veff:1}
\end{align}
If one considers the two-body interaction alone, that is, $q_0 = 0$, $\chi_0 = 0$, $q_1 = 0$ and $\chi_1  = 0$, the effective potential can be reduced as $U(R_0) =(g_0+2\pi )/\,(4\pi R_0^2) + \,g_1^2/\,(16\pi^2 \Omega^2 R_0^6)$, which is the same as discussed in ref.~\cite{Adhikari2004,Abdullaev2003,Saito2003}.
 The frequency of the small oscillations around the minimum of the effective potential $R^4_{min}=-3\,g_1^2/\,(4\pi(2\pi+g_0) \Omega^2)$ is given by
 \begin{align}
 \Omega^2_{br}=\frac{8(2\pi+g_0)^2\Omega^2}{3g_1^2}. \label{Jac13}
 \end{align}
A similar equation was obtained before by Saito and Ueda~\cite{Saito2003}.

We first analyse the stability of the condensate in the absence of three-body interaction from Eq.~(\ref{eq:veff:1}). 
\begin{figure}[!ht]
\centering\includegraphics[width=0.8\linewidth]{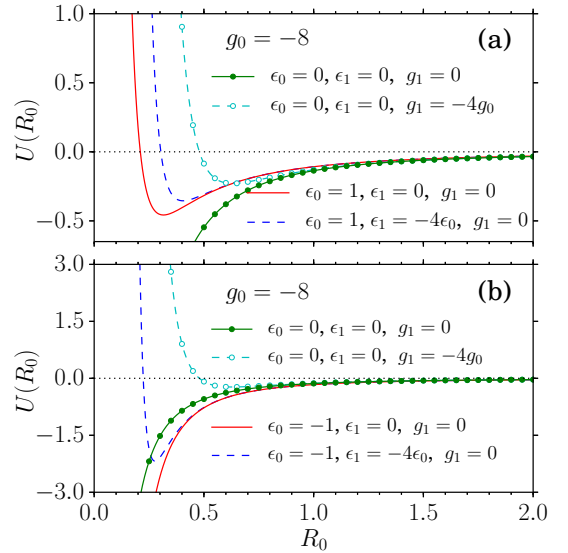}
\caption{(Color online) Plot of the effective potential $U(R_0)$ as a function of $R_0$ given in Eq.~(\ref{eq:veff:1}). 
(a)  $q_0>0$ and (b) $q_0<0$: Solid-red line represents the case of constant quantum fluctuation and constant two-body interaction, and dashed-blue line corresponds to the case with constant plus oscillatory quantum fluctuation and constant of two-body interaction. 
The solid-green line with filled circles shows the potential energy curve with constant two-body interaction alone, dashed-cyan line with empty circles denotes the curve for both constant and oscillatory two-body interaction.
}
\label{f1}
\end{figure}
Figs.~\ref{f1}(a) and \ref{f1}(b), show the effective potential as a function of $R_0$ for $\chi_0= \chi_1  = 0$, and $q_0>0$ and $<0$, respectively. We also show the potential energy curve with $q_0=0$ for comparison. It may be noted that, in two- and three dimensions, the trapless BECs with two-body interaction either collapse or expand due to the imbalance between the repulsive kinetic pressure and the attractive interaction~\cite{Saito2003}. When $q_0=0$, the trapless BEC in 2D can be stabilized by the oscillatory two-body interaction~\cite{Adhikari2004,Saito2003}. Also, it is worthwhile to note that one can stabilize the trapless two-component system by considering linear spin-orbit coupling~\cite{BAM_SOC}. However, in single component case, in the absence of oscillatory parts, it becomes unstable and the system expands to infinity. This is captured by the minimum of the dashed cyan line with empty circles in Fig.~\ref{f1}(a), while there is no minimum for the solid green line with filled circles. For a condensate of size $R$, the kinetic energy is proportional to $R^{-2}$ whereas the attraction is proportional to $R^{-d}$ in $d$ dimensions, and an effective potential for $R$ is the sum of these energies. The effective potential, therefore, has a minimum only for $d=1$. On the other hand, if we include a suitable quantum fluctuation then the system becomes stable without the presence of an oscillatory two-body interaction ($g_1=0$). In this case, the two-body attraction can be balanced by the repulsive force due to the quantum fluctuation. But, it is unstable when both two-body and quantum fluctuations are attractive, $g_0 < 0$ and $q_0<0$, and is shown in Fig.~\ref{f1}(b) by dashed blue curve. Moreover, if we consider repulsive $q_1>0$ with attractive $g_0$ and $q_0$, then it becomes stable and is illustrated by the solid cyan curve in Fig.~\ref{f1}(b).
 
\subsection{Interplay of three-body and quantum fluctuation on the stability of BECs with two-body interaction}

In this section, we discuss the interplay of three-body and quantum fluctuation on the stability of trapless BECs in the absence of oscillating nonlinearity. We have used $g_0<0$, $\chi_0>0$ and $q=0, \pm 1$.
\begin{figure}[!ht]
\centering\includegraphics[width=0.8\linewidth]{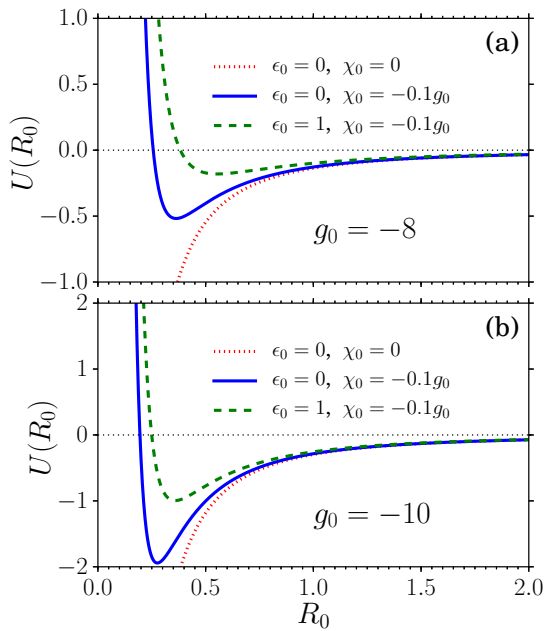}
\caption{Plot of the effective potential $U(R_0)$ of Eq.~(\ref{eq:veff:0}) as a function of $R_0$ with the effect of the three-body interaction quantum fluctuations for (a) $g_0 =-8$ and (b) $g_0 =-10$. 
}
\label{f3}
\end{figure}
In Fig.~\ref{f3}, the potentials are shown for the case of trapless condensate in the (i) absence of both three-body and quantum fluctuations (doted-red line), (ii) presence of three-body and absence of quantum fluctuation (solid-blue line), and (iii) presence of both three-body and quantum fluctuations (dashed-green line). Fig.~\ref{f3}(a)-(b) show the potential energy curves for two different values of two body interaction strength, say, $g_0=-8$ and $g_0=-10$. It is easy to see that the depth of the potential well increases with $g_0$ along with the three-body interactions and quantum fluctuations. 

\subsection{Effect of temporal modification of scattering length}
Next, we consider both constant as well as oscillatory nonlinearity for the two- and three-body interactions and quantum fluctuations.
\begin{figure}[!ht]
\centering\includegraphics[width=0.8\linewidth]{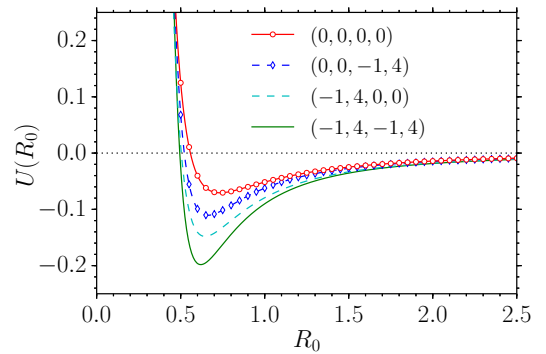}
\caption{Plot of the effective potential $U(R_0)$ of Eq.~(\ref{eq:veff:0}) as a function of $R_0$. Here $g_0 = -8$, $g_1 = -4 g_0$ and $(q_0, q_1, \chi_0, \chi_1)$ values are shown in the legends.}
\label{f5}
\end{figure}
The stability of trapless BECs with two-body interaction for constant (slowly varying) and oscillatory (rapidly varying) part has been explored~\cite{Adhikari2004,Abdullaev2003,Saito2003}. 
However, the effect on the inclusion of both constant and oscillatory parts of three-body interaction and quantum fluctuations has not been studied in trapless BECs in 2D. Hence, in the present study, we are also interested in analyzing the effect of both the constant and the oscillatory parts of the quantum fluctuations and three-body interactions on the stability of trapless BECs in 2D. Moreover, trapless BECs with constant and oscillatory parts of the two- and three-body interaction have been discussed in 3D~\cite{Sabari2010}. 

In Fig.~\ref{f5}, we show the potential energy curves of the trapless system for four different cases:  (i) constant and oscillatory two-body interaction alone, (ii) constant and oscillatory two- and three-body interactions, (iii) constant and oscillatory two-body and quantum fluctuations and (iv) constant and oscillatory  two-, three-body interactions and quantum fluctuations. From Fig.~\ref{f1}(a), we realize that one can stabilize the trapless BECs in 2D by tuning the strength of the quantum fluctuations without a trap and also in the absence of an oscillatory nonlinearity. Also, from Figs.~\ref{f3}, it is possible to stabilize the 2D trapless BEC without any oscillatory nonlinearities. So, from variational results, we come to know that the three-body interactions and quantum fluctuations play an important role in the stability of the system in 2D free space.

\section{Numerical Results}

\label{sec3}
To confirm the above analytical results, we solve the time-dependent GP equation (\ref{Jac2}) numerically through the split-step Crank-Nicholson method \cite{Muruganandam2009}. 
There could be many ways of stabilizing a trapless system numerically. In the course of time evolution of the GP equation certain initial conditions are necessary for the stabilization of a trapless system with a specific nonlinearity above a
critical value~\cite{Saito2003,Adhikari2002}.
If the size of the condensate is close to the desired size, then the condensate gets stabilized at large times. This procedure could also be followed in an experimental attempt to stabilize a system.
To solve the GP equation for large nonlinearity $\vert g(t) \vert$, $\vert q(t) \vert$ and $\vert \chi(t) \vert$, one may start with the Thomas-Fermi approximation for the wave function obtained by setting all the derivatives in the GP equation to zero, which is a good approximation for large nonlinearity \cite{Dalfovo1999,Muruganandam2009,Adhikari2002,Thogersen2009,Fetter2009}. Alternatively, the harmonic oscillator solution is also a good starting point for small values of nonlinearity as in this paper. The typical discretized space and time steps for solving this numerical method are $0.01$ and $0.0001$, respectively. 

In the numerical simulation, it is important to remove the harmonic trap while increasing the nonlinearity for obtaining the stability. Otherwise the oscillations that arise due to sudden removal of trap may lead to collapse due to attraction. In the course of time iteration, the coefficients of the nonlinear terms are increased from $0$ at each time step as 
\begin{align}
g(t) & = f(t) g_f  \{a_1-b_1\sin[\,\Omega\,\, (t-\tau)]\},\notag  \\
\chi(t) & = f(t) \chi_f \{a_2-b_2\sin[\,\Omega\,\, (t-\tau)]\} , \notag  \\
q(t) & = f(t) q_f \{a_3-b_3\sin[\, \Omega\,\, (t-\tau)]\}, \label{eq:num:sol}
\end{align} 
where $f(t)=t/\tau$ for $0 \leq t \leq \tau$ and $f(t)=1$ for $t > \tau$. At the same time the trap is removed by changing $d(t)$ from $1$ to $0$ by $d(t)=1-f(t)$. 
During this process, the harmonic trap is removed, and after the $g_f$, $\chi_f$ and $q_f$ are attained at time $\tau$, the periodically oscillating nonlinearity $g(t)=g_f[a_1-b_1 \sin \Omega t  ]$, $\chi(t)=\chi_f[a_2-b_2 \sin \Omega t  ]$ and $q(t)=q_f[a_3-b_3 \sin \Omega t ]$ are effected for $t> \tau$~\cite{Sabari2010,Adhikari2004,Saito2003,PSEKAR}. The correct performance of the above numerical scheme is important for the stabilization. If the attraction due to the final value of  nonlinearity after switching off the external trap is strong for the size of the condensate, then the system becomes highly attractive in the final stage and it eventually collapses. If the attraction due to the final value of  nonlinearity after switching off the external trap is weak for its size, the system becomes weakly attractive in the final stage and it expands to infinity~\cite{Adhikari2004}.

\begin{figure}[!ht]
\begin{center}
\includegraphics[width=0.99\linewidth]{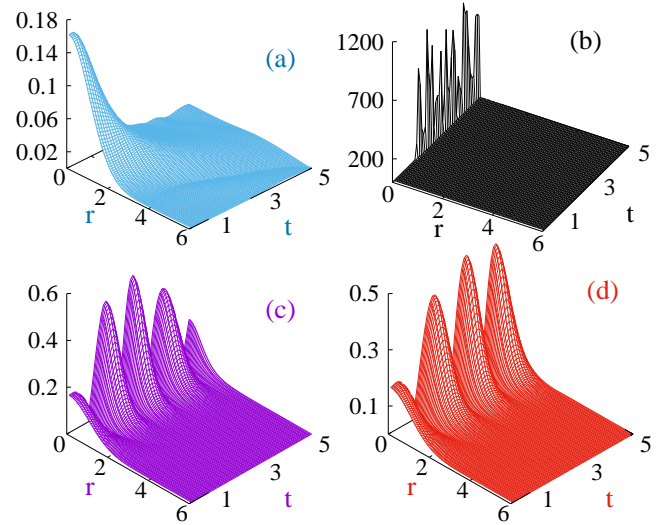}
\end{center}
\caption{Stabilization of 2D trapless, attractive BECs by repulsive three-body interaction and quantum fluctuations. (a) $g_f=-5$ and $a_1=1$ (b) $g_f=-15$ and $a_1=1$, (c) $g_f=-15$, $\chi_f=-0.2*g_f$ and $a_1=1$, $a_2=1$, and (d) $g_f=-15$, $q_f=0.5$ and $a_1=1$, $a_3=1$, and all other parameters are set to zero in all four cases.}
\label{f6}
\end{figure}
First, we study the stability of trapless BECs in the absence of oscillatory forces, $g_1=q_1= \chi_1  = 0$. As we discussed in the previous section, in the case of trapless BECs with two-body interaction alone, the repulsive kinetic pressure and the attractive interaction force cannot balance, and the condensate always collapses or expands~\cite{Adhikari2004,Abdullaev2003,Saito2003,Sabari2010}. The space-time plot of the density $\vert \phi(r,t)\vert ^2$ in Fig.~\ref{f6} illustrates this expansion and collapse of the system for $g_0=-5$ (panel (a)) and $g_0=-15$ (panel (b)), respectively.
However, as reported by Gammal et al~\cite{Gammal2000}, the inclusion of a repulsive three-body interaction can stabilize the trapless system in 2D free space. This case is prooved numerically in Fig.~\ref{f6}(c). Moreover, in Fig.~\ref{f6}(d), we show the stabilization of the trapless BEC by quantum fluctuations ($g_f=-15$, $q_f=0.5$ and $a_1=1$, $a_3=1$, and all other parameters are set to zero).
It means that one can stabilize the attractive trapless BECs in 2D by repulsive three-body interaction and also by quantum fluctuations. 

\begin{figure}[!ht]
\begin{center}
\includegraphics[width=0.85\linewidth]{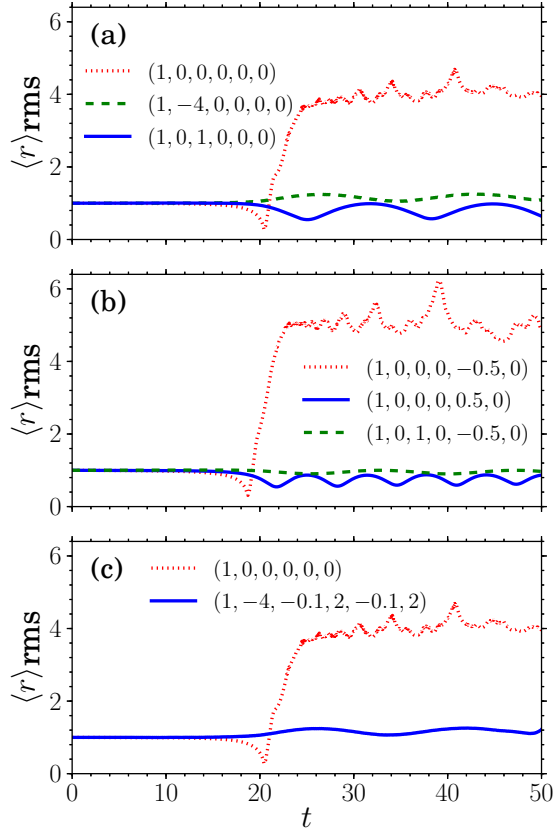}
\end{center}
\caption{Dynamics of the trapless BECs for different cases by means of the root mean squared distance $\langle r \rangle_{\mbox{rms}}$ as a function of time $t$. The parameters are $g_f = -20$, $\chi_f = 1$, and $q_f = 1$, and $(a_1, b_1, a_2, b_2, a_3, b_3)$ values are shown in the legends.}
\label{f12}
\end{figure}

Furthermore, in Fig.~\ref{f12} and in Table~\ref{table1}, we compare our results with previous results of Saito and Ueda~\cite{Saito2003}. 
\begin{table}[!ht]
\caption{Summary of BECs stability for $a_1 = 1$. See Eq.~(\ref{eq:num:sol}) for more details on the parameters $a_i$, $b_i$, $i=1,2,3$.}       
\label{table1}
\begin{center}
\begin{tabular}{c|r|r|r|r|r|l}
\hline\hline
\multicolumn{1}{c|}{case}       & \multicolumn{1}{c|}{$b_1$} & \multicolumn{1}{c|}{$a_2$} 
   & \multicolumn{1}{c|}{$b_2$} & \multicolumn{1}{c|}{$a_3$} & \multicolumn{1}{c|}{$b_3$} 
   & \multicolumn{1}{l}{remarks}\\ 
\hline 
 1 &    $0.0$ &   $0.0$ & $0.0$ &  $0.0$ &   $0.0$ &     {unstable} \\
\hline
 2 &   $-4.0$ &   $0.0$ & $0.0$ &  $0.0$ &   $0.0$ &     {stable}   \cite{Saito2003}   \\
\hline
 3 &    $0.0$ &  $1.0$  & $0.0$ &  $0.0$ &   $0.0$ &     {stable}    \\
\hline
 4 &    $0.0$ &   $0.0$ & $0.0$ & $-0.5$ &   $0.0$ &     {unstable}    \\
\hline
 5 &    $0.0$ &   $0.0$ & $0.0$ &  $0.5$ &   $0.0$ &     {stable}    \\
\hline
 6 &    $0.0$ &  $1.0$  & $0.0$ & $-0.5$ &   $0.0$ &     {stable}    \\
\hline
 7 &   $-4.0$ & $-0.1$  &  $2.0$  & $-0.1$ &    $2.0$  &     {stable}    \\
\hline\hline
\end{tabular}
\end{center}
\end{table}
In Fig.~\ref{f12}(a), dashed-green and solid-blue curves show the stabilization of the system in the presence of both constant and oscillatory two-body interactions, and constant two- and three-body interactions, respectively. However, the system is unstable in the presence of constant two-body interaction alone as illustrated by dotted-red line in Fig.~\ref{f12}(a). 

In Fig.~\ref{f12}(b), we show the effect of the three-body interaction and interplay between three-body and quantum fluctuations. Here, in the absence of oscillating nonlinearity, we stabilized the trapless BECs with the aid of the constant part of either the three-body interactions or both three-body interactions and quantum fluctuations. 
\begin{figure}[!ht]
\begin{center}
\centering\includegraphics[width=0.49\linewidth]{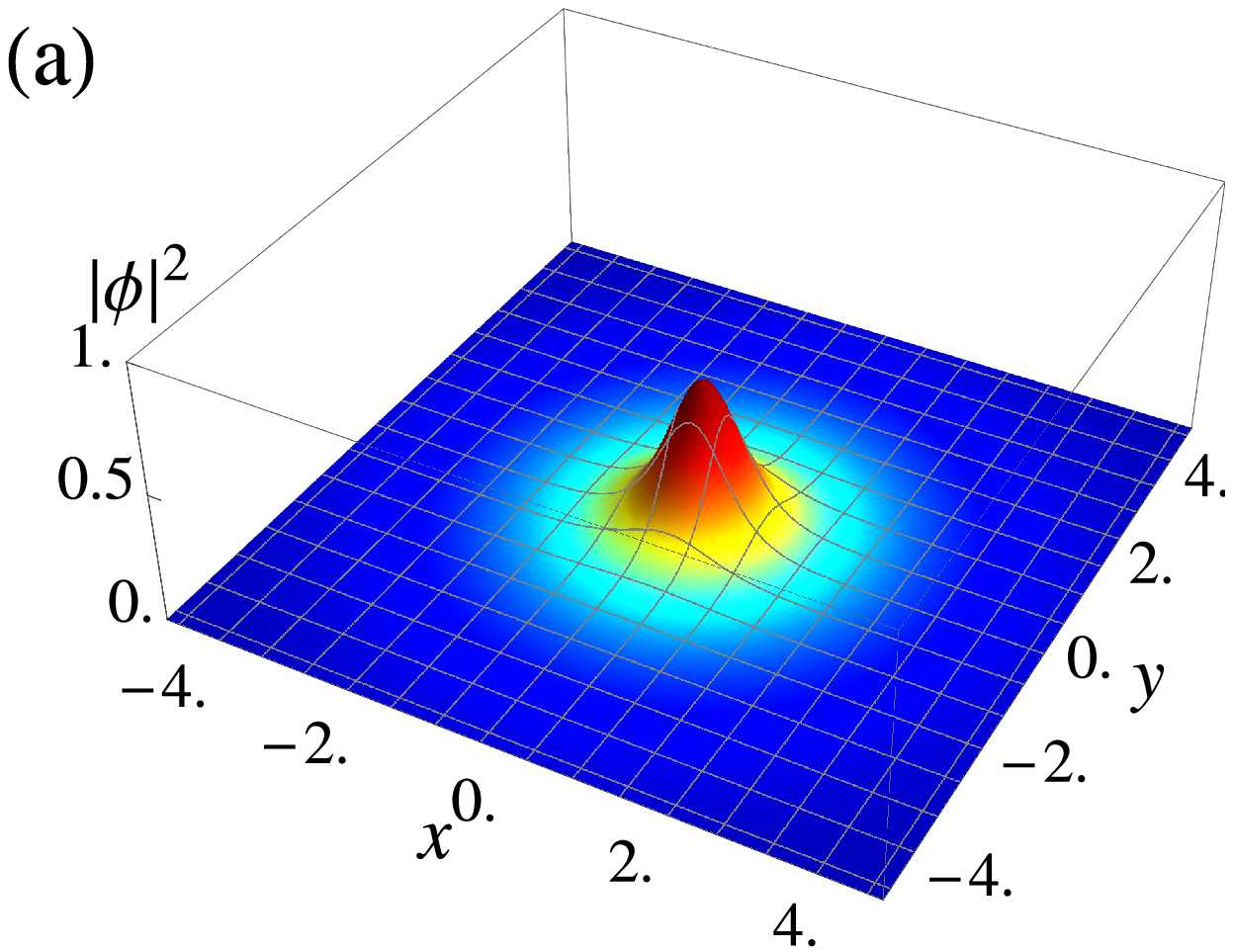}
\centering\includegraphics[width=0.49\linewidth]{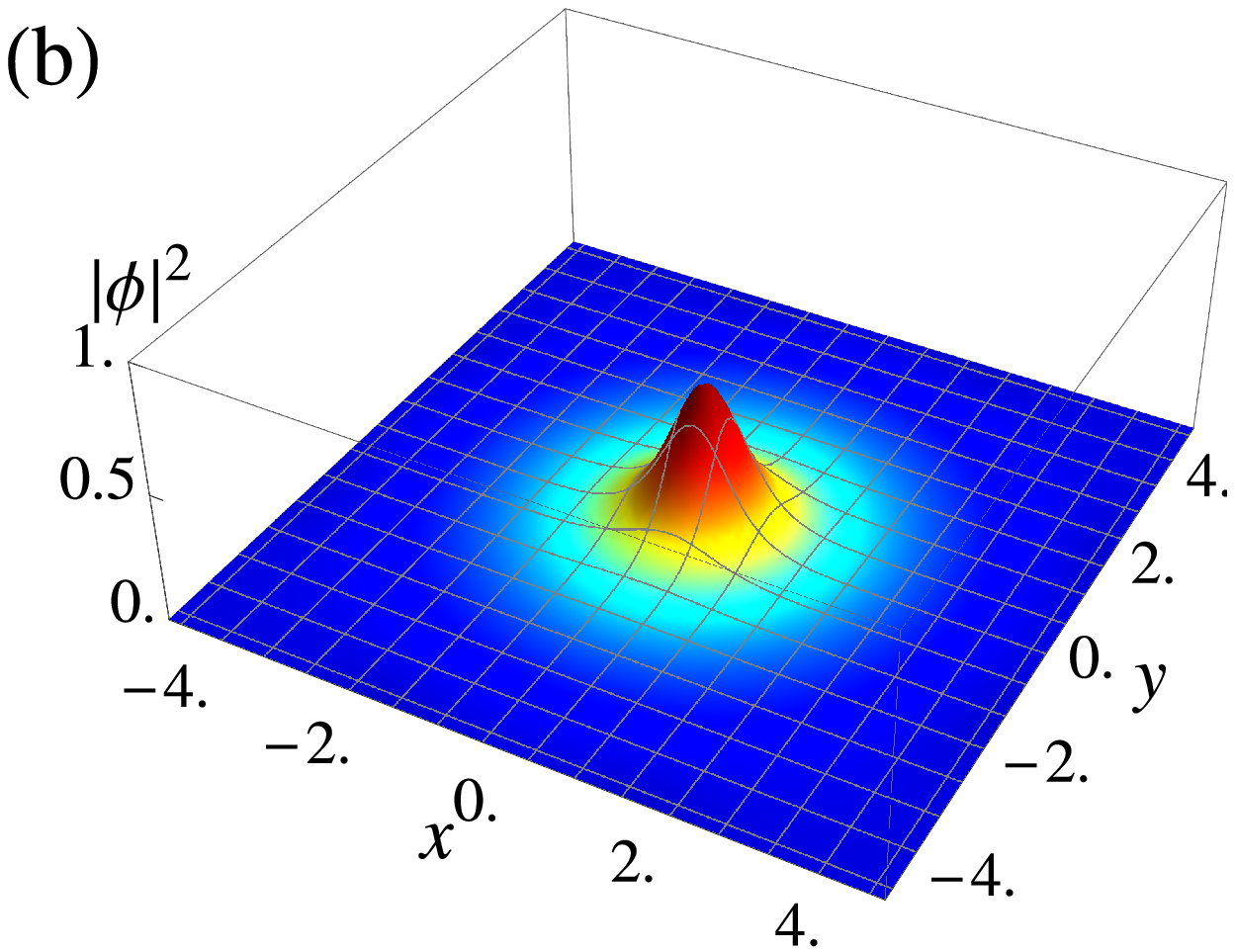}
\centering\includegraphics[width=0.49\linewidth]{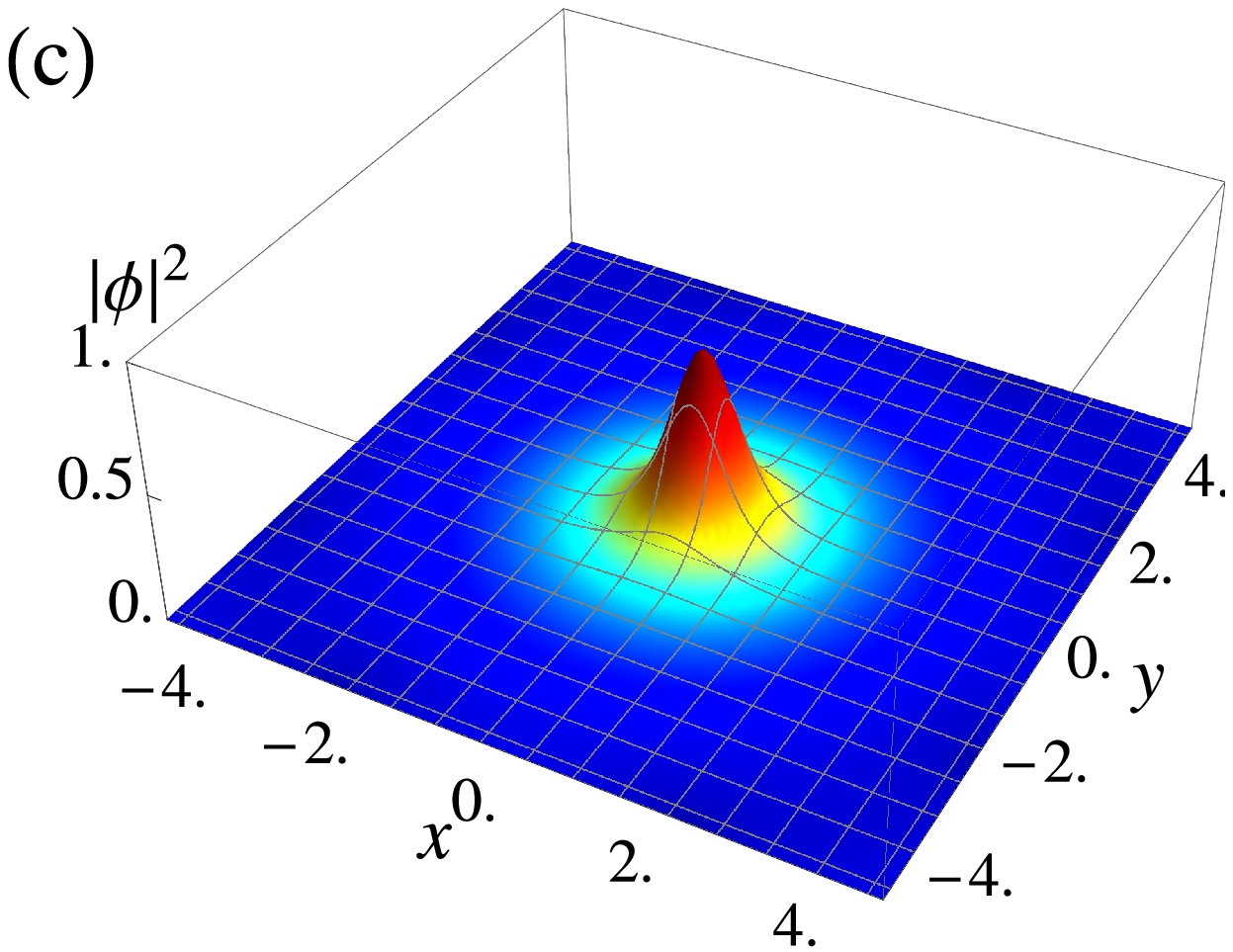}
\centering\includegraphics[width=0.49\linewidth]{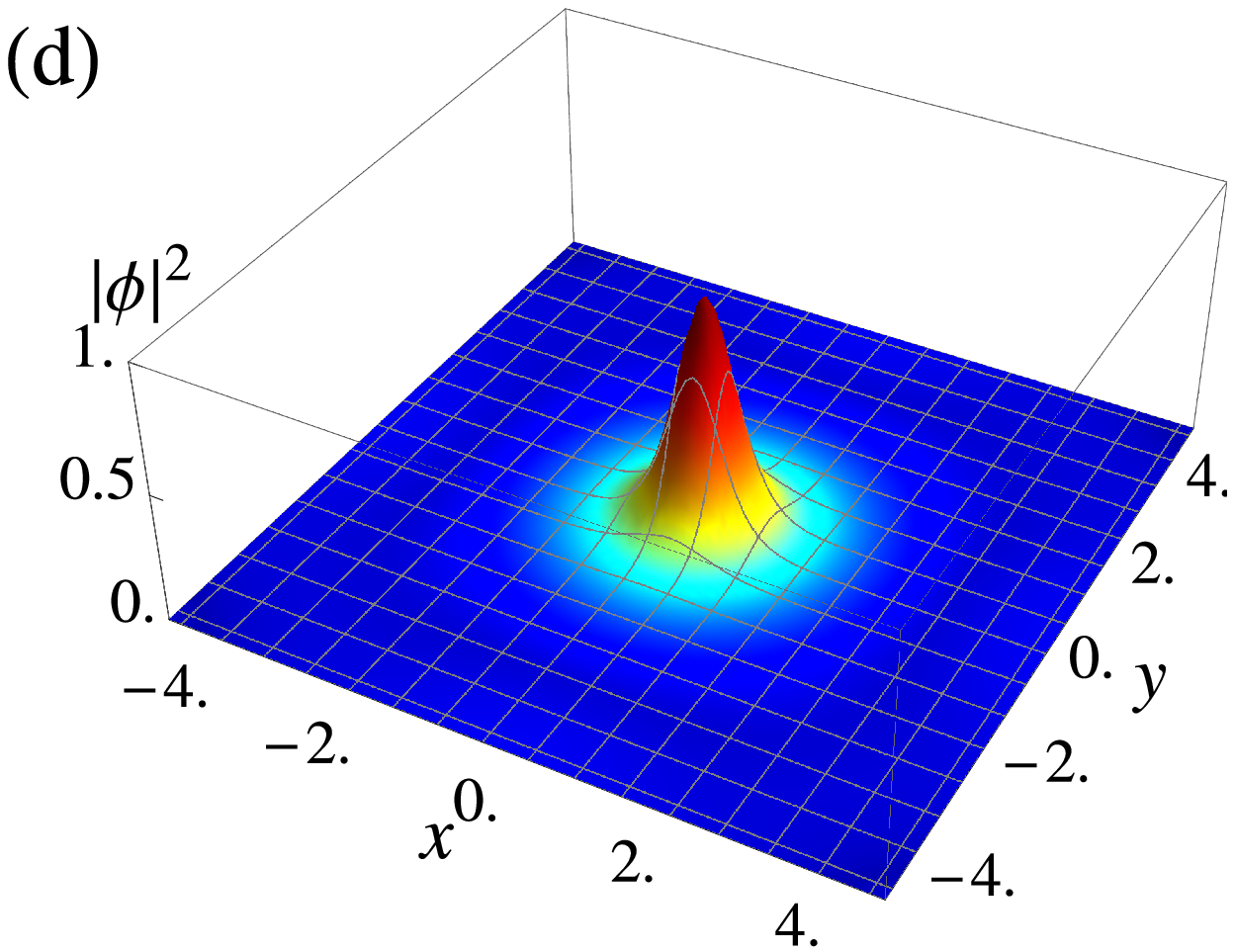}
\centering\includegraphics[width=0.49\linewidth]{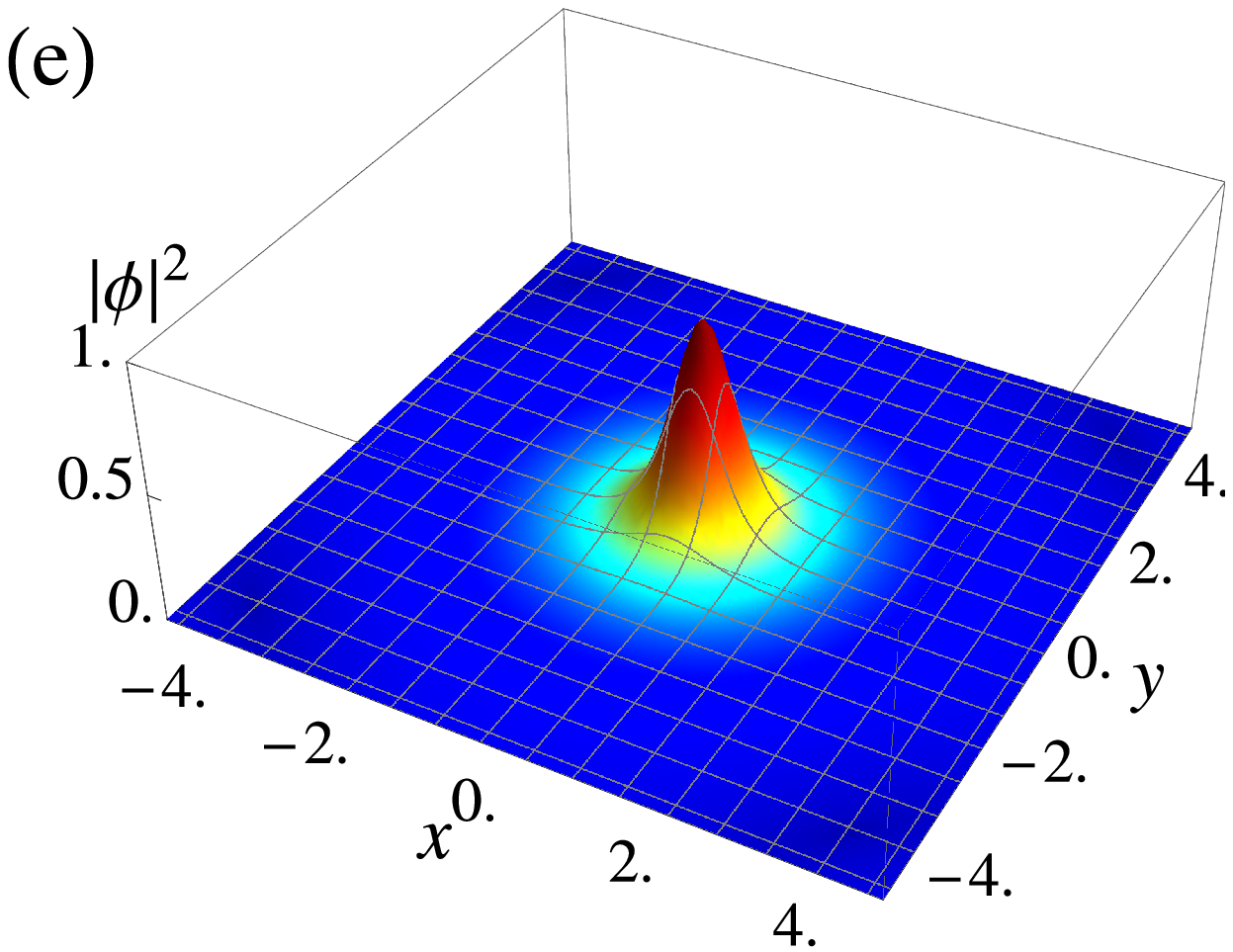}
\centering\includegraphics[width=0.49\linewidth]{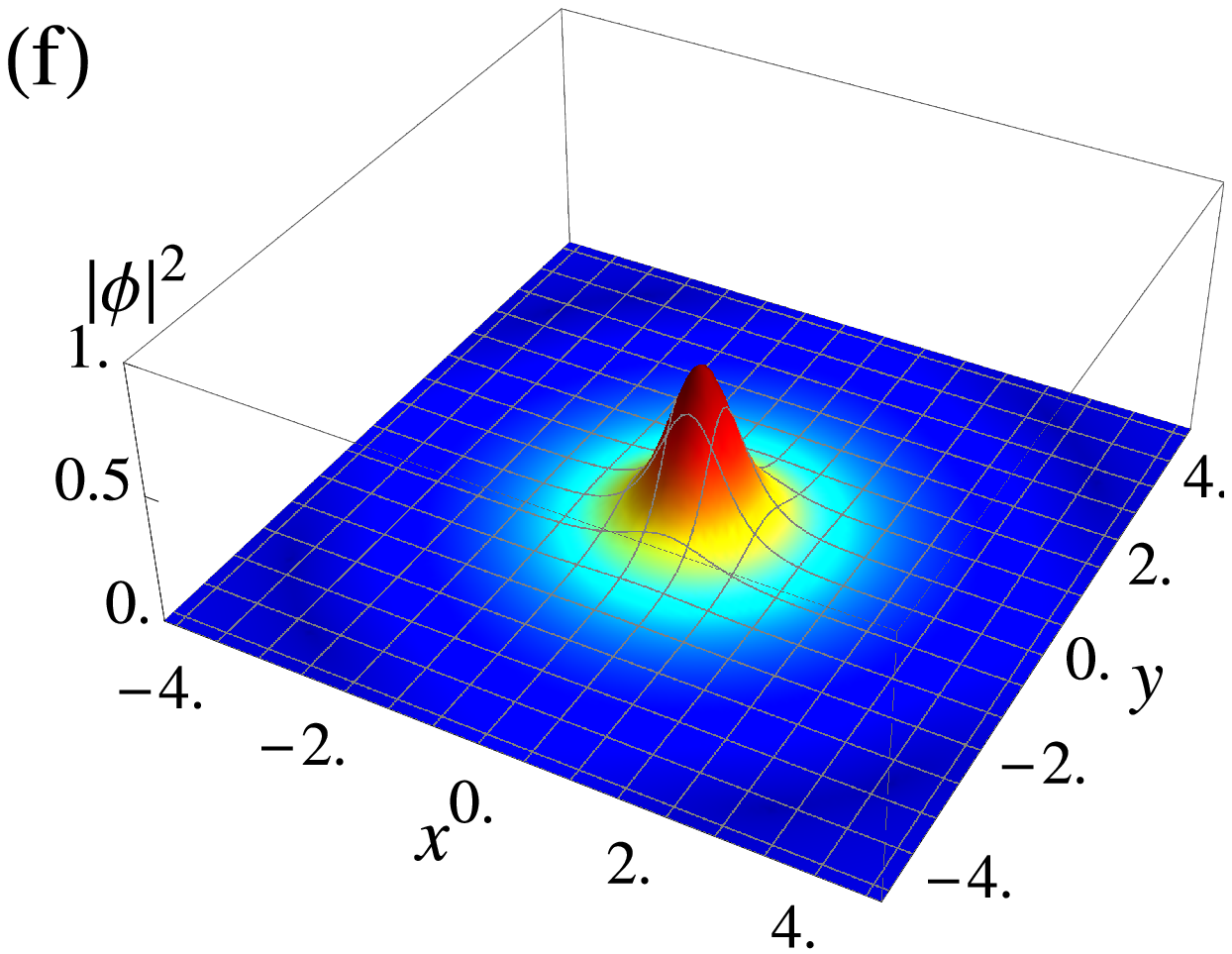}
\end{center}
\caption{Snapshots of the density $\vert \phi \vert^2$ for different instances of time after removing the trap with $\tau = 20$ as in Eq.~(\ref{eq:num:sol}): (a) $t = 20$, (b) $t=2 \tau$, (c) $t=3 \tau$, (d) $t=4 \tau$, (e) $t=5 \tau$ and (f) $t=6 \tau$.} 
\label{f13}
\end{figure}
Next, in Fig.~\ref{f12}(c), we have used both constant and oscillatory parts of all two- and three-body interactions along with quantum fluctuations, which also stabilizes the trapless BEC. In Table~\ref{table1}, we show the summary of BECs stability in free space in 2D. Fig.~\ref{f13} illustrates the stabilization of trapless BECs by quantum fluctuations by showing the snapshots of the two dimensional density taken at different instances of time. Further, the analytical solutions discussed in the previous section are verified through numerical simulations and the numerical results are consistent with the analytical  predictions. This confirms that, in the absence of any oscillating nonlinearity, one can stabilize the trapless 2D BECs by adding a constant quantum fluctuation. In addition, the stability of the trapless BECs in 2D can be increased by considering the three-body interaction and quantum fluctuations.

\section{Conclusion}
\label{sec4}
In conclusion, we have studied the stabilization of trapless BECs in 2D using GP equations with two- and three body interactions and in the presence of quantum fluctuations. Before studying the importance of three-body interaction and quantum fluctuations in terms of stabilization, we have performed variational analysis and derived the equation of motion to investigate the stability of trapless BECs. 
Most importantly, we have shown that the trapless BECs in two-dimensions can be stabilzed by including suitable quantum fluctuations without any oscillatory nonlinearities. Also, we have studied the effect of the addition of three-body interaction and quantum fluctuations with two-body interaction, which increases the stability of the trapless BECs in 2D. 
Moreover, based on the analytical results, we have studied the stability of 2D trapless BECs with rapid periodic temporal modulation of scattering length by using a Feshbach resonance. We have shown all possible ways of stabilizing trapless BECs in 2D in the presence of three-body interaction and quantum fluctuations.

Furthermore, we have analyzed different cases of interactions with presence or absence of constant or oscillatory three-body interactions and quantum fluctuations. We also verified our analytical results with numerical simulation using SSCN method. The numerical results confirm the analytical findings obtained by variational approach. From our analytical and numerical results, it is clear that one can stabilize and increase the stability of the trapless BECs in 2D by the inclusion of three-body interaction and quantum fluctuations.

\vskip 0.5 cm
{\bf Acknowledgement}
S.S. thanks the University Grants Commission (UGC) for offering support through a Post-Doc under the Dr. D.S. Kothari Post Doctoral Fellowship  Scheme (No.F.4-2/2006 (BSR)/PH/14-15/0046) and also the Department of Science and Technology-Science and Engineering Research Board (DST-SERB) for partial support (Grant No. PDF/2016/004106). 
K.P. acknowledges support from the National Board for Higher Mathematics (NBHM), DST-SERB, Indo-French Centre for the Promotion of Advanced Research IFCPAR (5104-2) and Council of Scientific and Industrial Research (CSIR), Government of India.
The work of P.M. forms a part of SERB, DST, Govt. of India sponsored research project (No. EMR/2014/000644).
\vskip 0.5 cm
{\bf Author contribution statement}
All authors contributed equally to the paper.

\end{document}